%% file: main.tex
\documentclass[fleqn,usenatbib]{mnras}
\pdfoutput=1
\usepackage{newtxtext,newtxmath}
\usepackage[T1]{fontenc}
\usepackage{ae,aecompl}

\usepackage{graphicx}	% Including figure files
\usepackage{amsmath}	% Advanced maths commands
\usepackage{amssymb}	% Extra maths symbols
\usepackage{url}
\usepackage{siunitx}
\title[ELT-scale AO RTC on Intel Xeon Phi]{ELT-scale Adaptive Optics real-time control with the\\
         \space Intel Xeon Phi Many Integrated Core Architecture}

\author[D. R. Jenkins et al.]{
David R. Jenkins\thanks{E-mail: d.r.jenkins@durham.ac.uk},
Alastair Basden,
and Richard M. Myers.\\
CfAI, Department of Physics, Durham University, DH1 3LE, UK
}

\date{Accepted XXX. Received YYY; in original form ZZZ}

\pubyear{2018}
\begin{document}

\label{firstpage}
\pagerange{\pageref{firstpage}--\pageref{lastpage}}
\maketitle

\input{content/abstract}


%%%%%%%%%%%%%%%%%%%%%%%%%%%%%%%%%%%%%%%%%%%%%%%%%%

%%%%%%%%%%%%%%%%% BODY OF PAPER %%%%%%%%%%%%%%%%%%

\input{content/introduction}

\input{content/methods}

\input{content/discussion}

\section*{Acknowledgements}

Real-time AO work by the Durham group is supported by the EU H2020 funded GreenFlash project, ID 671662, under FETHPC-1-2014, the UK Science and Technology Facilities Council consolidated grant ST/P000541/1, and an STFC PhD studentship, award reference 1628730.

%%%%%%%%%%%%%%%%%%%%%%%%%%%%%%%%%%%%%%%%%%%%%%%%%%



\bibliographystyle{mnras}
\bibliography{refs} % if your bibtex file is called example.bib

%%%%%%%%%%%%%%%%%%%%%%%%%%%%%%%%%%%%%%%%%%%%%%%%%%

%%%%%%%%%%%%%%%%% APPENDICES %%%%%%%%%%%%%%%%%%%%%

% \appendix

% \input{content/appendixI}
% \input{content/appendixII}

%%%%%%%%%%%%%%%%%%%%%%%%%%%%%%%%%%%%%%%%%%%%%%%%%%

% Don't change these lines
\bsp	% typesetting comment
\label{lastpage}
\end{document}

%% file: content/abstract.tex
%!TEX root = ../main.tex

\begin{abstract}
We propose a solution to the increased computational demands of Extremely Large Telescope (ELT) scale adaptive optics (AO) real-time control with the Intel Xeon Phi Knights Landing (KNL) Many Integrated Core (MIC) Architecture. The computational demands of an AO real-time controller (RTC) scale with the fourth power of telescope diameter and so the next generation ELTs require orders of magnitude more processing power for the RTC pipeline than existing systems. The Xeon Phi contains a large number $(\geqslant64)$ of low power x86 CPU cores and high bandwidth memory integrated into a single socketed server CPU package. The increased parallelism and memory bandwidth are crucial to providing the performance for reconstructing wavefronts with the required precision for ELT scale AO. Here, we demonstrate that the Xeon Phi KNL is capable of performing ELT scale single conjugate AO real-time control computation at over \SI{1.0}{\kilo\hertz} with less than \SI{20}{\micro\second} RMS jitter. We have also shown that with a wavefront sensor camera attached the KNL can process the real-time control loop at up to \SI{966}{\hertz}, the maximum frame-rate of the camera, with jitter remaining below \SI{20}{\micro\second} RMS. Future studies will involve exploring the use of a cluster of Xeon Phis for the real-time control of the MCAO and MOAO regimes of AO. We find that the Xeon Phi is highly suitable for ELT AO real time control.
\end{abstract}

%% file: content/introduction.tex
%!TEX root = ../main.tex

\section{Introduction}

Ground based optical and near-infrared astronomical telescopes suffer image degradation due to the optical aberrations introduced by the Earth's turbulent atmosphere. Adaptive optics \citep[AO,][]{Babcock1953} is a technique that has been successfully deployed to help reduce the effect that the atmosphere has on scientific observations. The basic operation of AO involves detecting the shape of an incoming wavefront using a wave-front sensor (WFS) and using this information to correct for the atmospheric perturbations using an optical correcting element, usually a deformable mirror (DM). The wavefront measurements are normally approximated by measuring the local wavefront slope at a number of discreet points in the pupil plane and the approximate wavefront phase can be reconstructed from these local slope measurements using a typically computationally intensive approach.

The atmosphere is constantly changing, and so adaptive optics systems need to operate at rates consistent with the rate of change of the atmospheric conditions. This is characterised by the atmospheric coherence time, $\tau_0$, typically of order \SI{1}{\milli\second} at optical wavelengths of around \SI{500}{\nano\meter}.  Not only should the exposure time of the WFS be $<\tau_0$ to capture a 'snapshot' of the turbulent phase, but the correction also needs to be applied within this time window otherwise the turbulent wavefront will have evolved past the point of measurement by $\geq$\SI{1}{\radian} of wavefront error and the correction after this time will no longer be valid.

\subsection{AO systems}

\begin{figure}
  \includegraphics[width=\columnwidth]{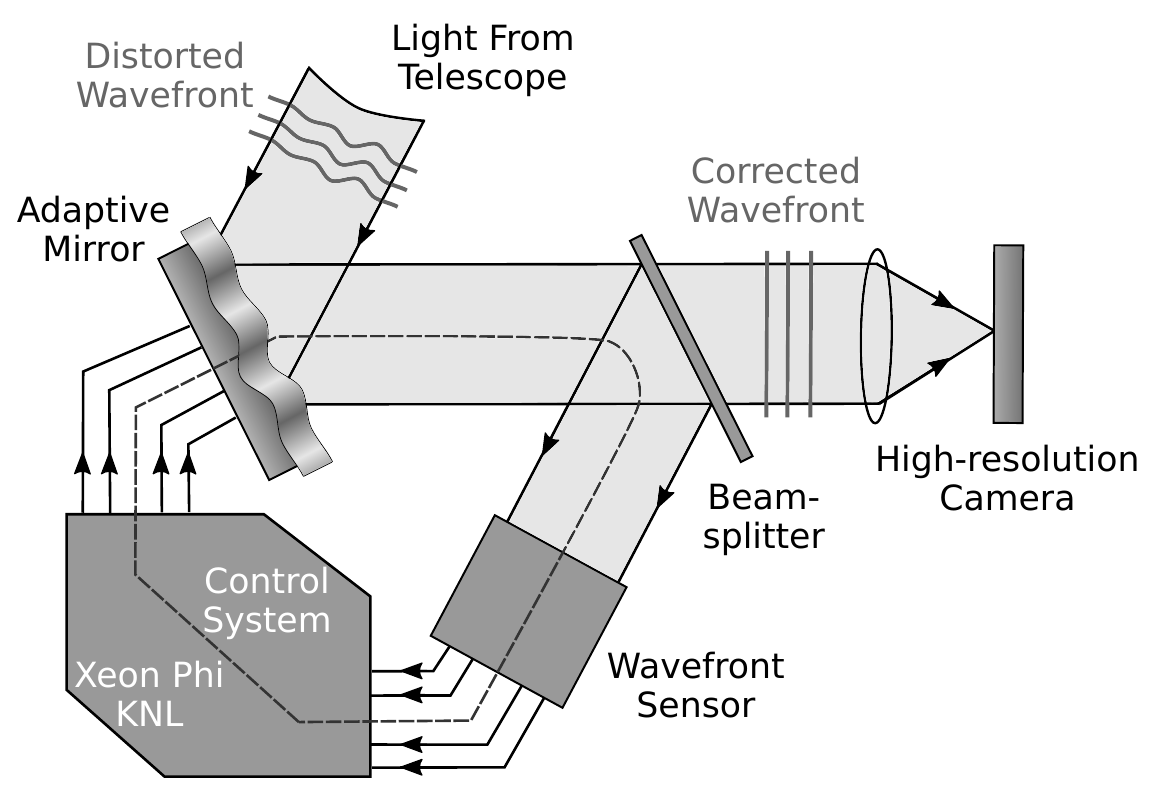}
  \caption{Schematic overview of closed loop Adaptive Optics. The control system is responsible for taking WFS images, processing them, reconstructing the incident wavefront and delivering the corrections via the adaptive mirror.}
  \label{fig:ao_overview}
  \end{figure}

An adaptive optics Real-Time Controller (RTC) is the hardware and software responsible for ensuring that the AO correction is computed and applied at the required rate for the atmospheric conditions (Figure \ref{fig:ao_overview}), typically of order \SI{1}{\kilo\hertz} for visible wavelengths, such that correction is applied within an atmospheric coherence time. The RTC takes wavefront sensor information (images) as input, calculates local wavefront slopes, computes the residual wavefront shape, and converts this to the required commands to be sent to the DM. The required high frame rate places strict requirements on the performance of the RTC so that it can meet the demands set out by the requirements of the AO instrument. Additionally, the variation in frame rate (the jitter) must also be low, otherwise scientific performance begins to suffer. An AO system's requirements are mostly defined by the parameters that quantify the strength and conditions of the atmospheric turbulence present at the site of observation, by the telescope size, and by the AO system type. These parameters can change for different types of scientific observation and for different operating wavelengths.

There are different classifications of AO control systems ranging from comparatively simple Single Conjugate AO (SCAO), which uses a single guide star for correction with a single DM, to more complex systems such as Multiple Conjugate AO \citep[MCAO,][]{Dicke1975,Beckers1988,Johnston1994} and Laser Tomographic AO \citep[LTAO,][]{Murphy1991,Tallon1990} which use many different reference stars and can employ multiple correcting elements. This paper will concentrate on the SCAO regime.

\subsubsection{ELT-scale AO RTC}

The dependence of AO system requirements on telescope diameter is an extremely important consideration for the next generation of Extremely Large Telescopes (ELTs), including the Giant Magellan Telescope \citep[GMT,][]{Johns2004}, the Thirty Meter Telescope \citep[TMT,][]{Stepp2004} and the European Southern Observatory \citep[ELT,][]{Spyromilio2008}, with primary mirror diameters greater than \SI{20}{\meter}. The computational complexity of the conventional wavefront reconstruction algorithm scales with the fourth power of telescope diameter; this is due to the dimensions of the control matrix which are governed by the number of correcting elements in the DM and the number elements of the WFS which both scale to the second power of telescope diameter. This presents a huge challenge in the process of designing an RTC suitable for ELT scale AO, both in the choice of hardware suitable to process the computational demands and with producing software capable of delivering performance that meets the requirements of the AO system.

Many typical AO reconstruction techniques involve computing one or more large matrix-vector-multiplications (MVMs) where the dimensions of the matrix are defined by the number of degrees of freedom (DoF) in the system. A control matrix, which contains information relating to how certain wavefront measurements correspond to the appropriate actuator commands, is multiplied by a vector containing the wavefront slope measurements, the results of which can yield the wavefront shape required for correction. A control matrix of dimensions (NxM) is made up of N DoF of a wavefront slope measurement and M actuators in the correcting element. This makes the MVM very large for ELTs requiring on the order of $N^2$ calculations where $N\approx M$.

The large MVM calculation is a memory-bandwidth bound problem, since the processing time is dependent on the rate that the processing unit can read the large matrix from memory (typically several GB), each AO RTC cycle. This therefore favours computational architectures with large memory bandwidth. The Intel Xeon Phi \citep{Intel2017} is one such architecture, containing a \SI{16}{\giga\byte} MCDRAM package, with a measured bandwidth as high as \SI{480}{\giga\byte\per\second}.  Modern graphics processing units (GPUs) also have large memory bandwidths.  However, a standard x86 CPU will typically have an achievable memory bandwidth of <\SI{90}{\giga\byte\per\second} and is therefore at a massive disadvantage for rapid processing of large MVMs.  They are therefore not well suited for ELT-scale AO RTC.  The Xeon Phi contains a large number of cores, has a large memory bandwidth, and is also programmable as a conventional processor.

\subsection{Real-time Controller Hardware}

The first consideration when designing an AO RTC is to choose hardware that can meet the requirements of the AO system, both in terms of input and output (I/O) interfaces for the instruments and in terms of computational performance for the algorithms required. The computational requirements are largely dictated by the reconstruction problem size. Historically, a variety of hardware architectures have been used for AO RTC, including digital signal processors \citep[DSPs,][]{Fedrigo2006}, field programmable gate arrays \citep[FPGAs,][]{Fedrigo2006,RodriguezRamos2012}, central processing units \citep[CPUs,][]{Basden2010} and graphics processing units \citep[GPUs,][]{Basden2010,Truong2012}.

These architectures have proved capable for previous and current AO systems with varying advantages and disadvantages for each. The main disadvantage with DSPs, FPGAs and GPUs is the time cost associated with designing, writing and, if necessary, modifying the RTC software. The main advantage of FPGAs and DSPs is deterministic behaviour. Due to their more general computing nature, CPU based systems can be at a disadvantage when it comes to some specific computation problems, such as highly parallelisable problems which may be better suited for GPUs, however they are much easier to develop for, and are generally backwards compatible with common programming languages. For ELT-scale systems, two of the main challenges are scaling of these systems for the increased computational complexity (usually requiring many of these devices working in parallel), and future proofing the development of the system for updated hardware.

The Intel Xeon Phi processor family combines many low power x86 CPU cores utilising wide \SI{512}{\bit} vector registers with high bandwidth on-chip memory to enable acceleration of highly parallelisable tasks, while keeping the cores sufficiently fed with data. These x86 cores use the backwards compatible x86 instruction sets which are used in the vast majority of Intel and AMD based CPU systems. This allows the Xeon Phi to leverage the benefits both of having a CPU based architecture and of having a highly parallelisable workflow similar to that of GPUs. These attributes of the Xeon Phi make it a very interesting candidate for an ELT-scale AO RTC as they can be developed using conventional CPU programming techniques. However due to the relatively low performance of an individual Xeon Phi CPU core, properly utilising vectorisation and parallelism is essential for good performance.

In this paper we present an investigation of the suitability of the Intel Xeon Phi Many Integrated Core (MIC) architecture as the processor for ELT-scale AO RTC. Section \ref{sec:devAOxeonphi} summarises our efforts with adapting existing RTC software for use with the Xeon Phi platform, section \ref{sec:resultsdiscuss} presents our results and section \ref{sec:conclusion} concludes with a summary and a discussion of our work.

\subsubsection{Xeon Phi: Knights Landing}

\begin{figure}
\includegraphics[width=\columnwidth]{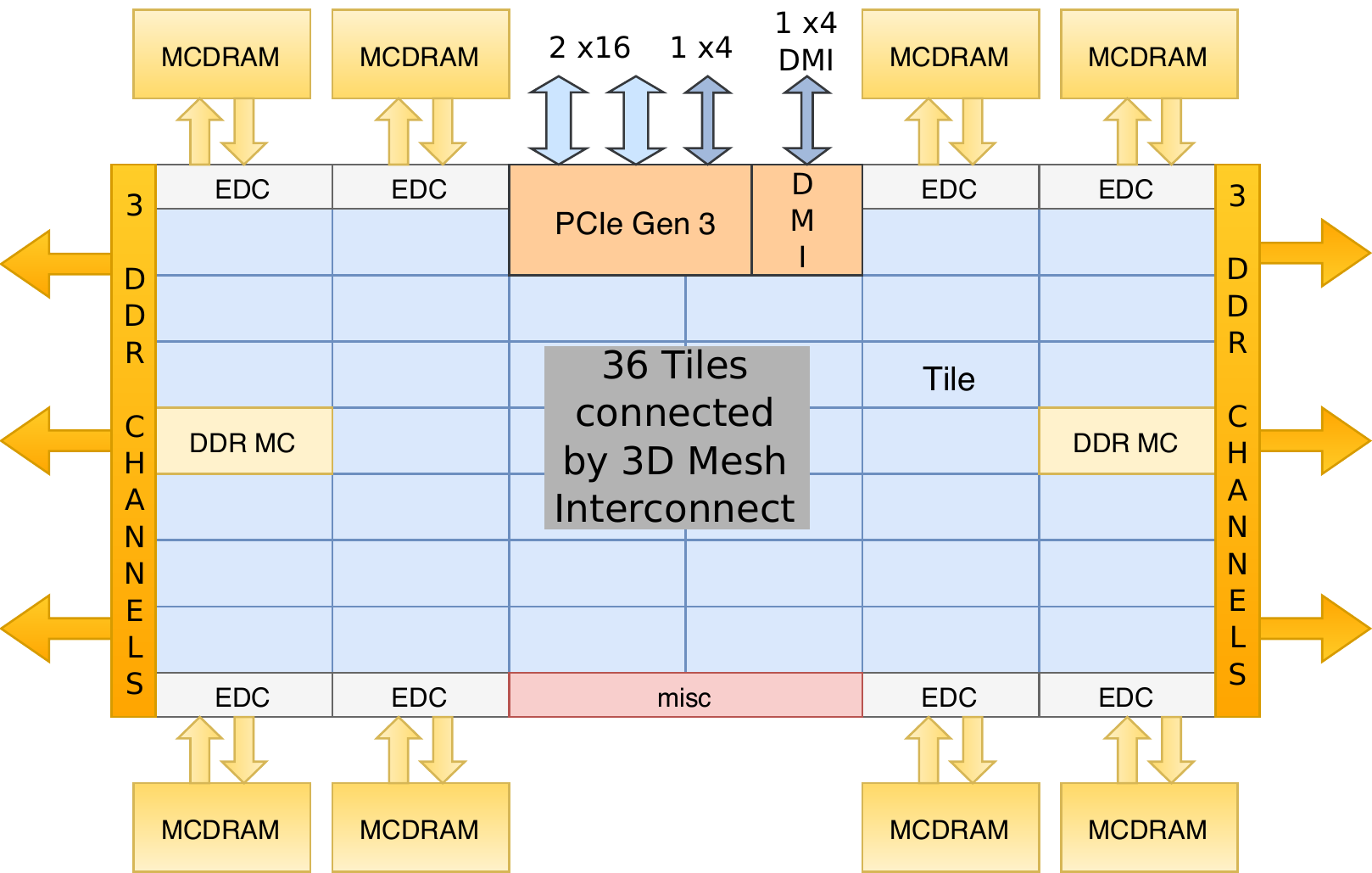}
\caption{Schematic of Xeon Phi Knights Landing CPU showing the MIC architecture along with the high bandwidth MCDRAM. Each tile contains two CPU cores, two vector processing units per core and a shared 1MB of Level2 cache. (DDR MC = DDR memory controller, DMI = Direct Media Interface, EDC = MCDRAM controllers, MCDRAM = Multi-Channel DRAM \citep{Intel2016})}
\label{fig:knlschema}
\end{figure}

Knights Landing (KNL) is the third generation Xeon Phi processor and is the first to be released in the self-booting socketed form factor, with a number of variants, as given in table~\ref{tab:knlspec}. Previous generation Xeon Phi chips were available as accelerators only. The KNL processor can therefore be used just like a conventional server processor and can run the Linux operating system and standard software environment. Existing applications can be ported to the Xeon Phi quickly: recompilation is usually not even required, though code will not be well optimised in this case.

The KNL introduces additional instruction sets, which can be utilised for improved performance, including \SI{512}{\bit} CPU registers, which allow Single Instruction Multiple Data (SIMD) operation on 16 single-precision floating point numbers simultaneously per core, per instruction cycle. This improves the parallelisation advantage of the KNL architecture over previous processors, which included a maximum of \SI{256}{\bit} wide vector registers. This \SI{512}{\bit} register is also included in forthcoming (and most recent) standard Intel CPUs, so any code optimisations made for this feature will also be applicable to future non-Xeon-Phi CPUs. However, it should be noted that for the KNL system, high utilisation of 512 bit instructions reduces the base core clock speed by 200Hz, which is taken into consideration in the peak SP TFLOPS calculated in Table~\ref{tab:knlspec}.

The wavefront sensor cameras can be directly attached to the KNL via the PCIe bus.  This is an advantage over accelerator cards such as previous the generation Xeon Phi cards and GPUs, where, unless specific effort is taken (often requiring specific network cards, and low level device programming), image data must be transferred first to the CPU from the camera, and then out to the accelerator (essentially 2 PCIe transfers, with increased latency and jitter). The previous generation of the Xeon Phi (Knights Corner) has also been investigated for AO RTC \citep{Barr2015}, showing promise for future processor generations (i.e.\ the KNL).

\begin{table}
  \centering
  \caption{Available Knights Landing models and their key specifications. The peak single precision (SP) TFLOPS is a theoretical calculation resulting from the core count, the clock speed (-200MHz for 512-bit vector operation), the vector register size, the number of vector process units per core and the number of floating point operations per fused-multiply-add. e.g for the 7210 model: $64\times(1.3-0.2)\times512/32\times2\times2=5325$ GFLOPS. The memory bandwidth is that as measured using the STREAM triad benchmark.}
  \begin{tabular}{ cccccc }
    \hline
    KNL & Core & Base CPU & Peak & Memory & Memory \\
    Model & count & Clock Speed & SP & Speed & Bandwidth \\
    & & (\si{\giga\hertz}) & TFLOPS & (\si{\giga T\per\second}) & (\si{\giga\byte\per\second}) \\
    \hline
    7210 & 64 & 1.3 & 4.51 & 6.4 & 450 \\
    7230 & 64 & 1.3 & 4.51 & 7.2 & - \\
    7250 & 68 & 1.4 & 5.22 & 7.2 & 480 \\
    7290 & 72 & 1.5 & 5.99 & 7.2 & - \\
    \hline
  \end{tabular}
  \label{tab:knlspec}
\end{table}

%% file: content/methods.tex
%!TEX root = ../main.tex

\section{Development of AO RTC using a Xeon Phi} \label{sec:devAOxeonphi}

A large component of the time and effort required to design and produce an AO RTC stems from development of the control software. For technologies such as DSPs, FPGAs and GPUs, this can be extremely time consuming and require specific technological expertise without any guarantee that the software will be in any way compatible with future devices. CPU program development is comparatively more straightforward with a choice of well documented and easily accessible programming languages to choose from which are compatible with a wide range of CPU based platforms.

\subsection{Best case performance for ELT-scale SCAO RTC}
In order to determine the best performance achievable with a Xeon Phi, we have developed a highly optimised algorithmic RTC, i.e.\ a simple software solution which performs all the necessary RTC algorithms using optimised library functions, but which is not pipelined, does not interact with real camera or DM hardware, and which is not user configurable. Therefore, although this RTC cannot be used in a real AO system, it gives some idea of the minimum frame computation time (or maximum frame rate) which can be achieved using this hardware. We note that an investigation using a full on-sky tested RTC is introduced in later sections.

\begin{figure}
    \includegraphics[width=\columnwidth]{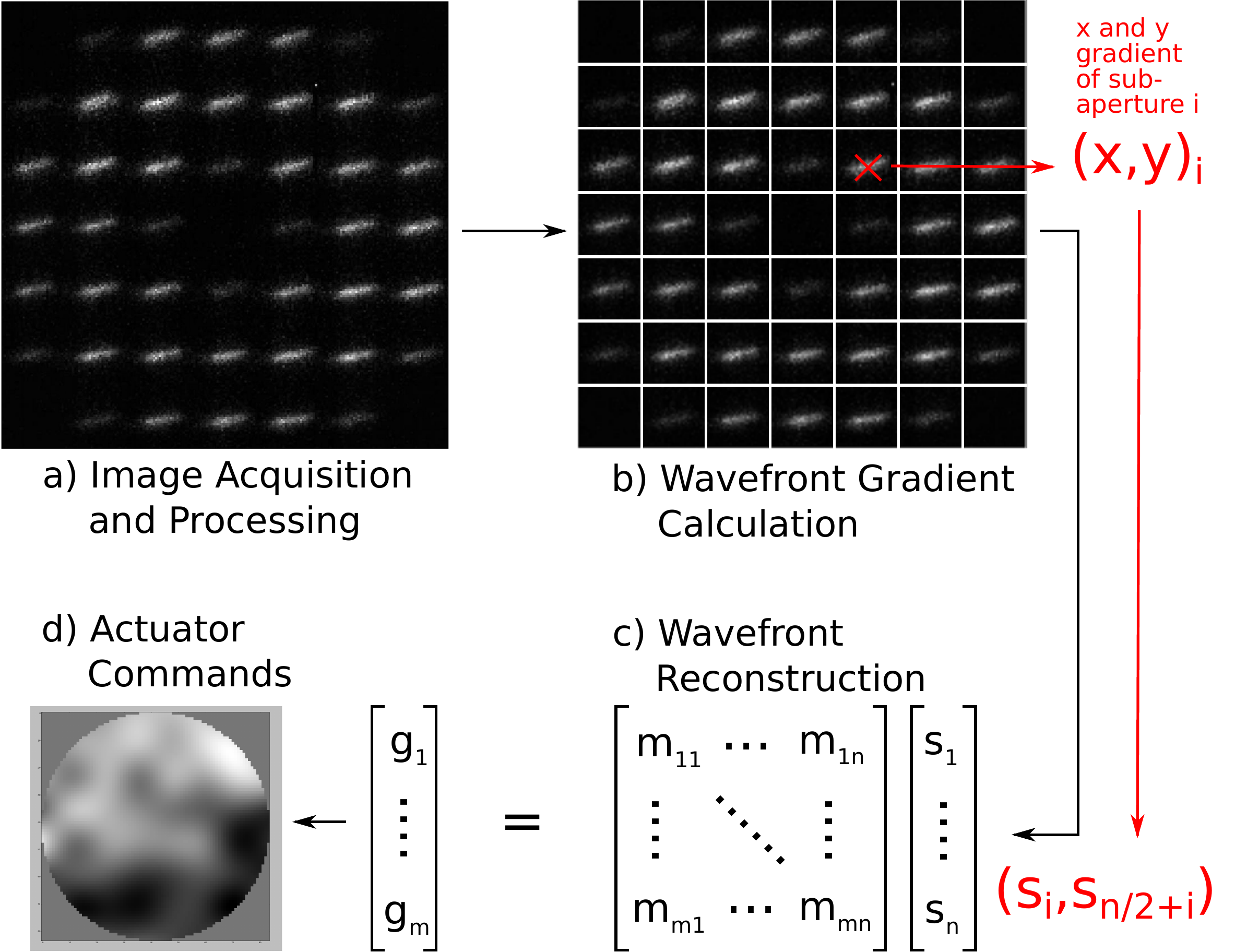}
    \caption{A figure showing the basic RTC operations, including a) image acquisition and processing (background subtraction, flat field application and threshold application), b) local wavefront gradient computation (using a centre of gravity algorithm), c) wavefront reconstruction (using a Matrix Vector Multiplication, MVM) and d) output of actuator commands. A thread will process a defined set of subapertures from beginning to end. For each subaperture, the local wavefront gradients are placed in a slope vector such that all the x gradients come first and then the y gradients ($(x,y)_i$ -> $(s_i,s_{n/2 +i})$). The result of the MVM is a vector of actuator commands which can be reformatted to show the resulting shape of the correcting element. }
    \label{fig:basicRTC}
\end{figure}

The simple simulator uses the OpenMP API \citep{openmp2015} for multithreading and performs pixel calibration on fake image data, centroiding of the calibrated pixels, an MVM reconstruction of the centroids and finally introduces a gain factor to the final result. The slope measurements are computed as if all pixels are available at once. This is the minimum computational requirement of an SCAO RTC and gives a base-line for best-case expected performance of the Xeon Phi.  Figure~\ref{fig:basicRTC} gives an overview of this system.

\subsection{Real RTC software using the Durham AO Real-time Controller}

We have optimised a real on-sky tested RTC for the Xeon Phi KNL in the form of The Durham Adaptive Optics Real Time Controller \citep[DARC,][]{Basden2010}. DARC is a freely available and on-sky proven AO RTC software package written in the C and Python programming languages. It is built upon a modular real-time core which allows it to be extended for many different AO RTC scenarios such as for different AO regimes like SCAO and MOAO and allows individual algorithms such as pixel calibration and wavefront reconstruction to be replaced or modified. The modular design also allows it to interface with many different devices for wavefront sensing and wavefront correction \citep{Basden2016}, making it flexible enough to be used in almost any AO situation. Because DARC is built on the C and Python programming languages it can be compiled and run on many different systems including the socketed Xeon Phi, as well as x86, POWER8 \citep{basden2015} and ARM processors. Within DARC, wavefront sensor images are processed in parallel, with sub-apertures being processed as soon as enough pixels have arrived at the computer.  DARC uses a horizontal processing strategy which allocates a similar workload to each thread, with threads being responsible for processing of a sub-aperture from start to finish (including calibration, slope calculation and partial reconstruction).  This means that AO latency can be reduced, since by the time the last pixels arrive at the computer, the majority of the processing for that frame has already been completed.  Here, we consider the optimisation of DARC for use with the Xeon Phi architecture, and report on performance investigations.

\subsection{Main areas of DARC optimisation for the Xeon Phi}
As an x86 CPU the Xeon Phi shares many attributes with standard CPU hardware.  However, it is also very different from previous CPUs with its many $(\geq64)$ low power cores, its high bandwidth memory and the \SI{512}{\bit} wide vector registers for improved SIMD performance. While most software developed for standard CPU systems can be compiled and run on Xeon Phi hardware with no alterations, to make the most of the new features, some optimisations are needed to best utilise the available hardware, these include:

\begin{enumerate}
    \item thread synchronisation, to make efficient use of all cores
    \item memory access, to optimise for the fast memory
    \item vectorisation, to take advantage of the wide vector registers.
\end{enumerate}

\subsubsection{BIOS, Operating system and kernel setup}
The operating system (OS) installed on the Xeon Phi used in this paper is CentOS Linux 7.3 \citep{centos2001}. To obtain the best low latency and low jitter performance various changes have been made to the default settings of the BIOS, the operating system and the kernel. The main changes to the BIOS settings involve turning off Intel Hyper-threading, which allows more logical threads to execute concurrently on hardware cores.  Removing Hyper-threading allows each software thread to be pinned to a single hardware core and removes scheduling inefficiencies caused when cores switch between different Hyper-Threads.

Other BIOS settings include Xeon Phi specific settings which relate to how the CPU handles memory addressing, with information available online \citep{Intel2015}, and different modes which determine how the fast Multi-channel DRAM (MCDRAM) is allocated, either accessible like standard RAM, reserved for the OS as a large last level cache (LLC), or a mixture of the two; these modes are termed 'flat', 'cache', and 'hybrid' respectively.

OS and kernel setup refers to options such as isolating certain CPU cores so that the OS doesn't schedule any program to run on these cores without specific instruction, and other options relating to CPU interrupts and different power and performance modes.

During our testing, we have identified that best performance is achieved with the CPU set to Quadrant memory addressing mode, and the MCDRAM was set to 'flat' mode. In 'flat' mode, the MCDRAM is visible to the CPU on a separate NUMA node from the standard RAM and so this must be addressed either by explicitly allocating the memory in the program (using a NUMA library), or by executing the program on the specific NUMA node to make use of the fast MCDRAM.  In this report the MCDRAM was allocated by running software with the `numactl' command with the `--membind=nodes' option, ensuring that the entire RTC is allocated on this NUMA node.  On the Xeon Phi, the MCDRAM is 16~GB in size, which is sufficient to fit a whole ELT-scale RTC.

\subsubsection{Multi-threading and synchronisation}
Multi-core CPU systems have become the norm in recent years leading to DARC being developed using a multi-threaded real-time core with the POSIX \citep{theopengroup2016} pthread library. The main method of ensuring thread synchronisation has been by the use of pthread mutexes and condition variables. A mutex is a mutual-exclusion variable which allows threads to `lock' a certain section of code, preventing other threads from accessing these protected regions. If a thread calls the lock function on an unlocked mutex variable, then that thread will be allowed to  proceed and then the mutex becomes locked. Any other threads which attempt to lock this mutex will have to wait at the lock function until the mutex is unlocked.

If multiple threads are waiting at a mutex then they will proceed one by one as the mutex is repeatedly locked and unlocked by the preceding thread. A thread waiting at a mutex will generally be descheduled by the operating system scheduler and put to sleep, reducing power consumption and freeing up the hardware for other threads to be scheduled. This works well for low order multi-core systems with 2-16 CPU cores, as it allows for more threads than physical CPU cores and the simultaneous descheduling and rescheduling of these few threads when they are waiting at the same mutex has little overhead.

\begin{figure}
    \includegraphics[width=\columnwidth]{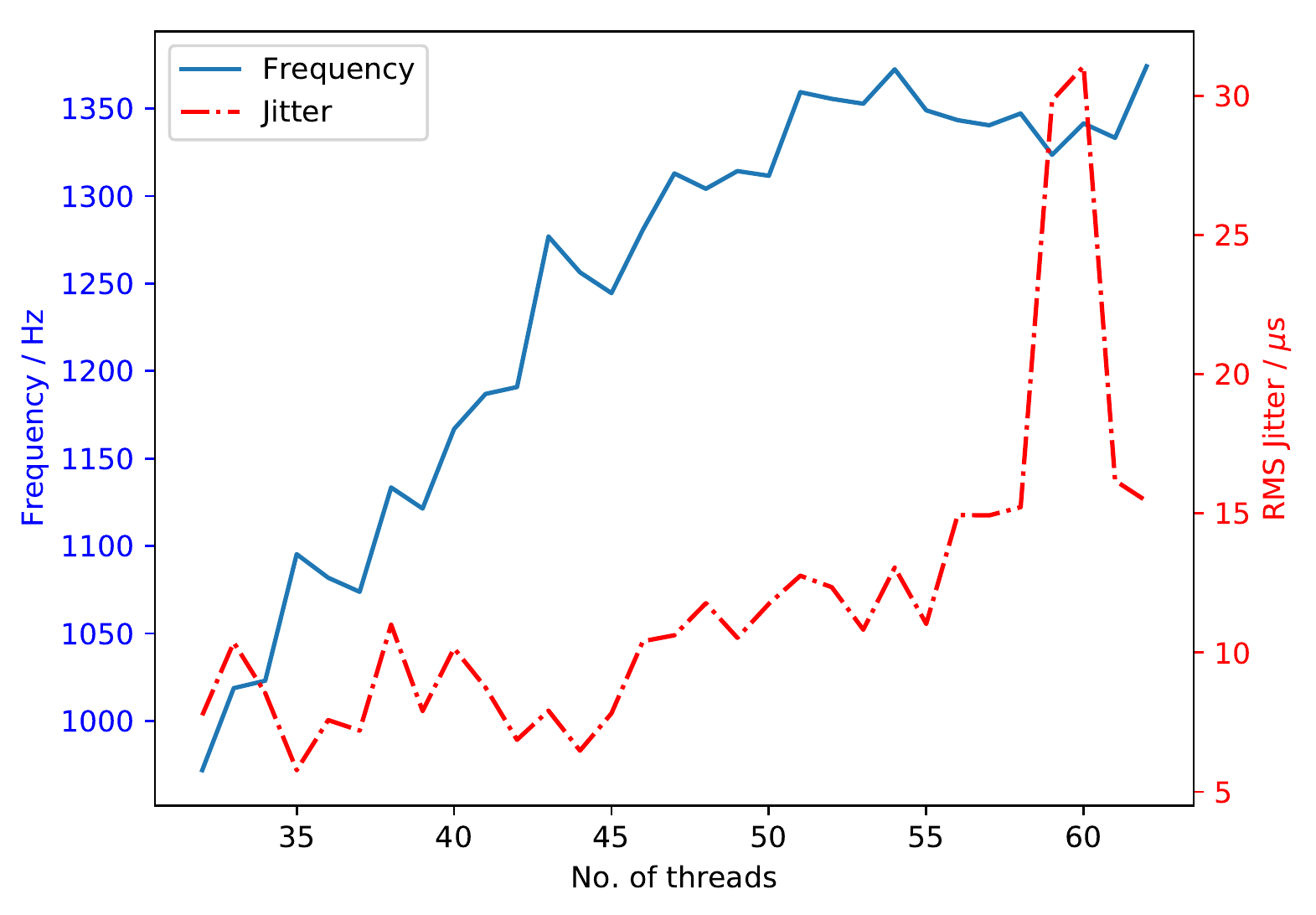}
    \caption{A measure of the average frequency of DARC running with different numbers of threads on Xeon Phi KNL, given in \si{\hertz}. Also shown is the RMS jitter of the frametime data used given in \si{\micro\second}. The periodic structure seen at higher thread counts is most likely due the thread allocation used to aid in vectorisation described in Section~\ref{sec:vectorisation}.}
    \label{fig:threadscaling}
    \end{figure}

However, for the Xeon Phi MIC architecture with $\geq64$ low power cores, DARC needs to be  configured to execute a single thread per core with >~48 threads to achieve maximum performance for ELT-scale SCAO  (Figure~\ref{fig:threadscaling}). This can cause problems when using mutexes and condition variables as the constant sleeping and waking of this large number of threads significantly increases latency. The solution that we have developed is to use a structure similar to mutexes called spinlocks, which also protect critical sections of parallel code but instead of sleeping and descheduling, threads simply wait until they can proceed.  This waiting process constantly consumes CPU cycles but increases the system's responsiveness which massively reduces the latency when using many threads as it is then much faster to resume operation.

Unfortunately, condition variables do not work with spinlocks and so we replace these where possible by simple volatile flag variables, taking care to ensure that thread safety is maintained.

\subsubsection{Vectorisation}\label{sec:vectorisation}
The \SI{512}{\bit} wide vector registers present on the Xeon Phi allow up to 16 single precision (SP) operations to be performed per cycle per CPU core. An operation in this case can be a fused-multiply add (FMA) operation which combines an addition and a multiplication, allowing up to 16 SP additions and 16 SP multiplications per instruction cycle. This is double the previous specification of \SI{256}{\bit} vector registers allowing a theoretical 2X speed up for vectorisable computations. Vectorisation is generally handled by the compiler: depending on the level of optimisation chosen at compile time, a certain amount of auto-vectorisation will occur.  However, steps can be taken to aid the compiler and investigate where vectorisation occurs or does not occur.  Essentially, if the compiler is able to detect that vector or matrix operations include 16-float boundaries at the same points, then these operations can be vectorised.  This therefore usually means that by aligning memory to the nearest 64~bytes, vectorisation will be aided.

The allocation of subapertures to specific threads within DARC can be optimised such that each thread processes a multiple of 16 slope measurements when calculating its own section of the wavefront reconstruction MVM. As each subaperture has 2 slope measurements, one for each of the x and y directions, we therefore ensure that the subapertures are allocated to threads such that each thread processes a multiple of 8 sub-apertures as a block.

Alignment of array memory to page cache boundaries is important so that the data required for the vectorised instructions can be loaded into the registers efficiently and with the right ordering. This can be done when allocating memory for the arrays using the posix \textunderscore memalign \citep{theopengroup2016} function call which aligns the amount of memory required at the specified boundary. The next step is to then ensure that sections which can be vectorised  are written in such a way that the compiler can apply auto-vectorisation; Intel provides a guide which details the necessary steps \citep{Intel2012}.

\subsubsection{Reducing the memory bandwidth requirement} \label{sec:halffloat}
Because the wavefront reconstruction MVM is memory bandwidth bound, due to the relatively simple mathematical operations but large data size, investigating ways to reduce the memory bandwidth dependence is an important consideration for ELT-scale AO RTC. A potential solution is to store the control matrix using \SI{16}{\bit} floating point format, rather than the conventional \SI{32}{\bit} format. The format used for the \SI{16}{\bit} floats is the IEEE 754 specification for binary16 \citep{IEEE754} which reduces the exponent from \SIrange[range-phrase=~to~]{8}{5}{\bit s} and the mantissa from \SIrange[range-phrase=~to~]{23}{10}{\bit s}.

This change does result in some loss of precision in the control matrix, however the available precision is still greater than that of the wavefront slope measurements (which are based on integer-valued detector measurements) and is therefore still considered sufficient for the reconstruction \citep{halfPrecisionFloat}. Every AO loop iteration, this control matrix is then loaded into CPU registers, converted to \SI{32}{\bit} format for operations (necessary since the Xeon Phi cannot perform \SI{16}{\bit} floating point mathematical operations), and the DM vector computed.  The reduction in memory bandwidth required can therefore reduce AO system latency.

\subsubsection{Optimising Parallel Vector-addition} \label{sec:optparvecadd}
As each DARC thread processes a set of subapertures from beginning to end, the result of each thread's execution is a partial DM vector for those subapertures. To combine these results into a final DM command vector they must be all be summed together. Previously DARC has achieved this by using a mutex to lock the final DM command vector whilst each thread adds its vector into it in turn. This works well for small numbers of threads on fast cores,  however for the KNL case where there are more threads on comparatively slower cores, this serial addition can be a source of a large amount of extra latency.

\begin{figure}
    \includegraphics[width=\columnwidth]{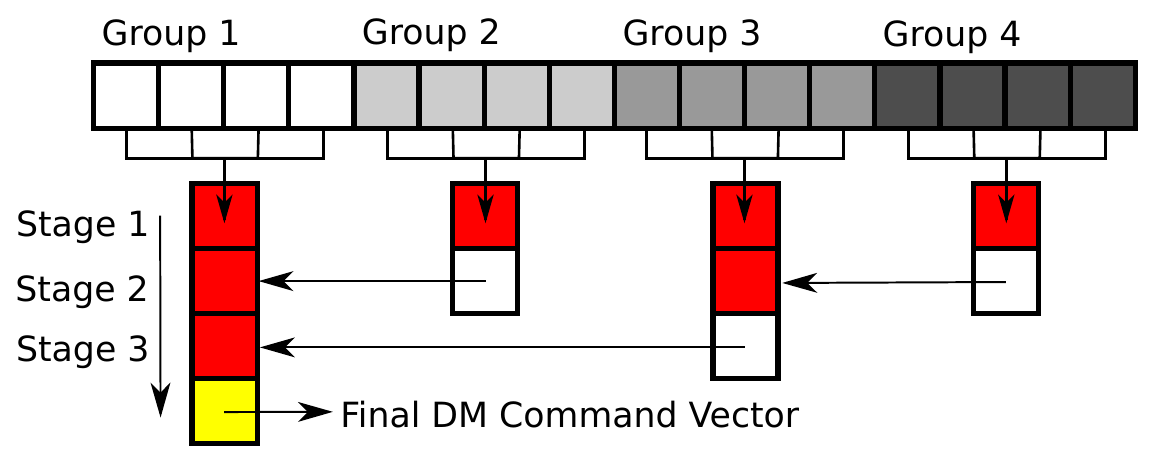}
    \caption{A schematic of the branching tree-like vector addition algorithm for a 4->2->2 situation with 16 threads, the first stage involves groups of four threads adding up their partial DM vectors, stages 2 and 3 reduce the resulting temporary DM commands to final DM command vector. This example allows up to 4 vector additions to happen in parallel and a total of 3 sequential stages instead of simply adding up all 16 threads' partial DM vectors sequentially. For larger thread counts, the effect is even more pronounced.}
    \label{fig:treeadd}
    \end{figure}

A solution that we have developed is a branching tree-like algorithm which allows groups of threads to add their partial DM vectors and so each group can work in parallel. A group will be defined by a spin-lock and a thread-barrier. Within the group a thread will get the lock whilst it copies its partial DM vector into a temporary output array and once each thread has copied it's partial vector it waits at the barrier for the other threads in the group to finish. At the completion of a groups work, one thread from each group will take ownership of the temporary output array and move on to the next group. This can be seen in Figure \ref{fig:treeadd} where each group adds its partial DM vector into the red box before that moves down to the next group where the process is repeated until the final DM command vector results.

\subsubsection{Reducing RTC using latency with asymmetric subaperture thread allocation} \label{sec:pipelining}

\begin{figure}
    \includegraphics[width=\columnwidth]{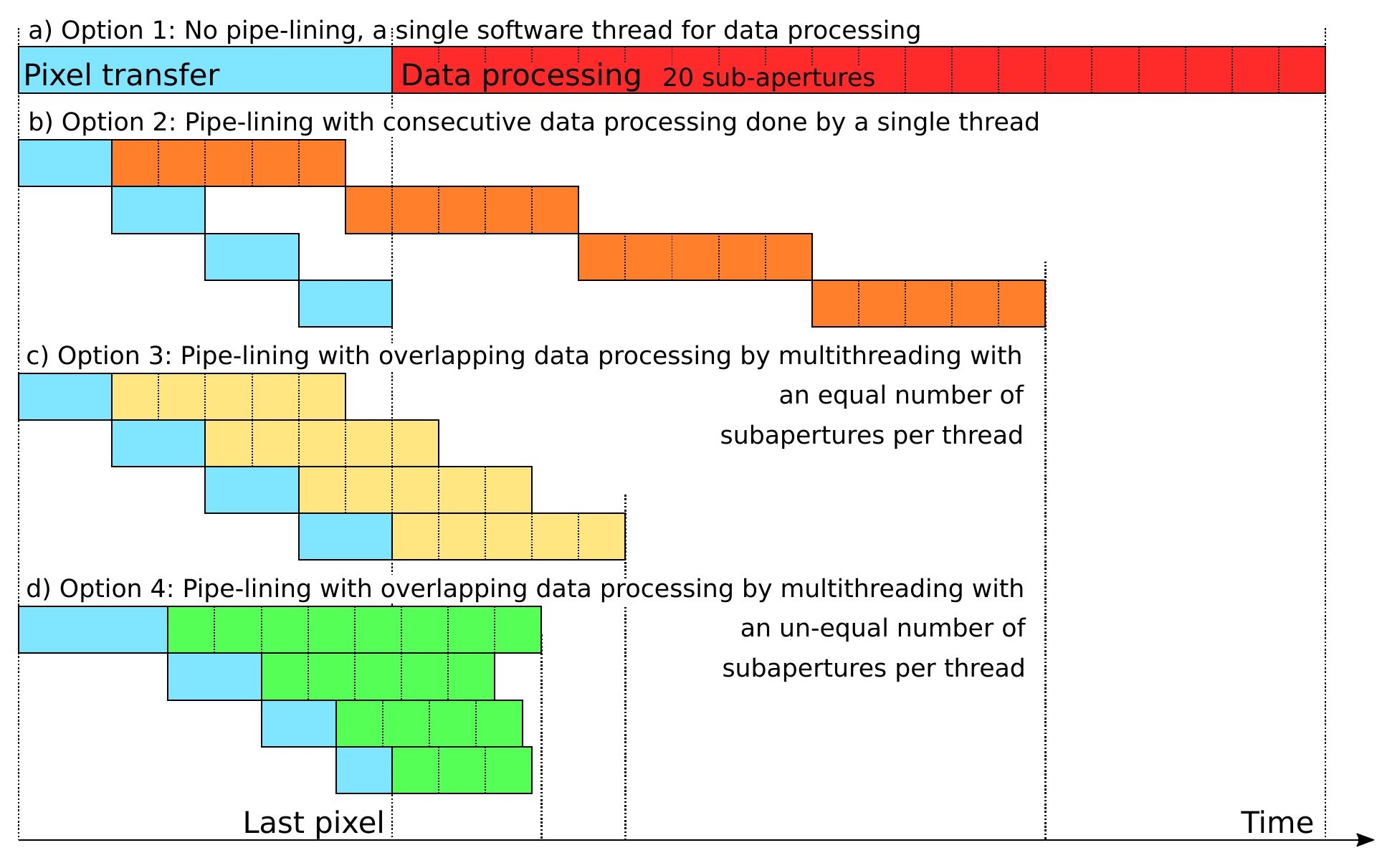}
    \caption{A comparison of the latency introduced via various pixel handling techniques. It shows that minimal latency is achieved via pipe-lining of the reconstruction using threads which process unequal numbers of subapertures such that they finish processing at roughly the same time.}
    \label{fig:pipelining}
    \end{figure}

The latency of an RTC is defined as the time from last pixel to readying the final DM command and so reducing this time interval is paramount to improving the performance of the RTC. The different options for handling and processing the pixel stream are shown in Figure~\ref{fig:pipelining}, with each subsequent option reducing the time taken from last pixel to end of thread computation.

As each DARC thread processes its subapertures from beginning to end, they must be allocated a specific set of subapertures to process, the most simple and naive way of assigning subapertures would be to divide them equally amongst the threads, Figure~\ref{fig:pipelining}(c). However as can be seen the threads which process the earlier arriving subapertures finish their work before the later threads, the subapertures are assigned equally among threads in an ascending order, as the thread number increases so does the time waiting for pixels, yet the processing time is constant.

Seen in Figure~\ref{fig:pipelining}(d) is an option whereby the subapertures are not allocated equally among the threads, the threads that process earlier subapertures are given more work to do and the later ones are given less. This ensures that the threads finish their work at roughly the same time and so helps to reduce the time between the last pixel arriving and the final DM command being sent out. However, the time waiting for pixels also changes as processing more subapertures requires waiting for more pixels, which can be seen in the different sized blocks for pixel waiting. This is a lot more complex in practice as there will not be a one-to-one relation between the number of sub-apertures and the number of pixels that need to be waited for.

%% file: content/discussion.tex
%!TEX root = ../main.tex

\section{Results and discussion} \label{sec:resultsdiscuss}
The performance of an AO RTC is generally defined by the time taken for it to process each frame, where a frame is a single image from a WFS and the processing involves calibrating the pixels, computing the centroids and reconstructing the wavefront before sending the results to a correcting element. There are different ways of determining the amount of time that it takes an RTC to process a frame and it depends on the definition of when a frame starts and when it ends. Timing for the entire AO RTC loop is generally taken from the time the first pixels arrive at the RTC core to the time when the last DM commands have been sent to the correcting element.

This encompasses the entire computation of the RTC especially when the main loop is pipelined, i.e., the processing is done for groups of pixels as they arrive due to the way the image sensors read-out the pixels.  However, in the case of a fast RTC and a slower camera, RTC latency will be artificially lengthened by periods waiting for camera pixels to arrive.  Here, we use the traditional definition of RTC latency, defined by the time taken between last camera pixels arriving at the RTC and last DM demand being computed.  This definition can therefore be computed entirely within the RTC hardware, though does not include delays due to the capture of camera pixels (e.g.\ by a frame grabber card, and transfer to the computer memory), or delays due to time taken for DM demands to leave the computer and arrive at the DM.  In addition to latency, we also report maximum stable RTC loop rate, i.e.\ the fastest rate at which the RTC can operate stably.

In cases where we present maximum RTC rate, i.e.\ without a camera attached, we define latency as the inverse frame time: in this case, the latency represents the minimum computation time for the RTC loop.

We also define the jitter of the AO RTC to be the variation in latency.  We present both rms jitter and also peak-to-peak (worst case) jitter.

\subsection{Suitability of Xeon Phi for ELT scale AO RTC} \label{sec:bestcasescao}
We configure the best case SCAO RTC simulator in an ELT configuration with $80 \times 80$ sub-apertures with a $0.25\times D$ central obscuration and $10 \times 10$ pixels per sub-aperture. This results in 4708 active sub-apertures, and therefore 9416 slope measurements. The number of DM actuators is 5170, based on a circular $81 \times 81$ actuator DM aperture.

\begin{figure}
\includegraphics[width=\columnwidth]{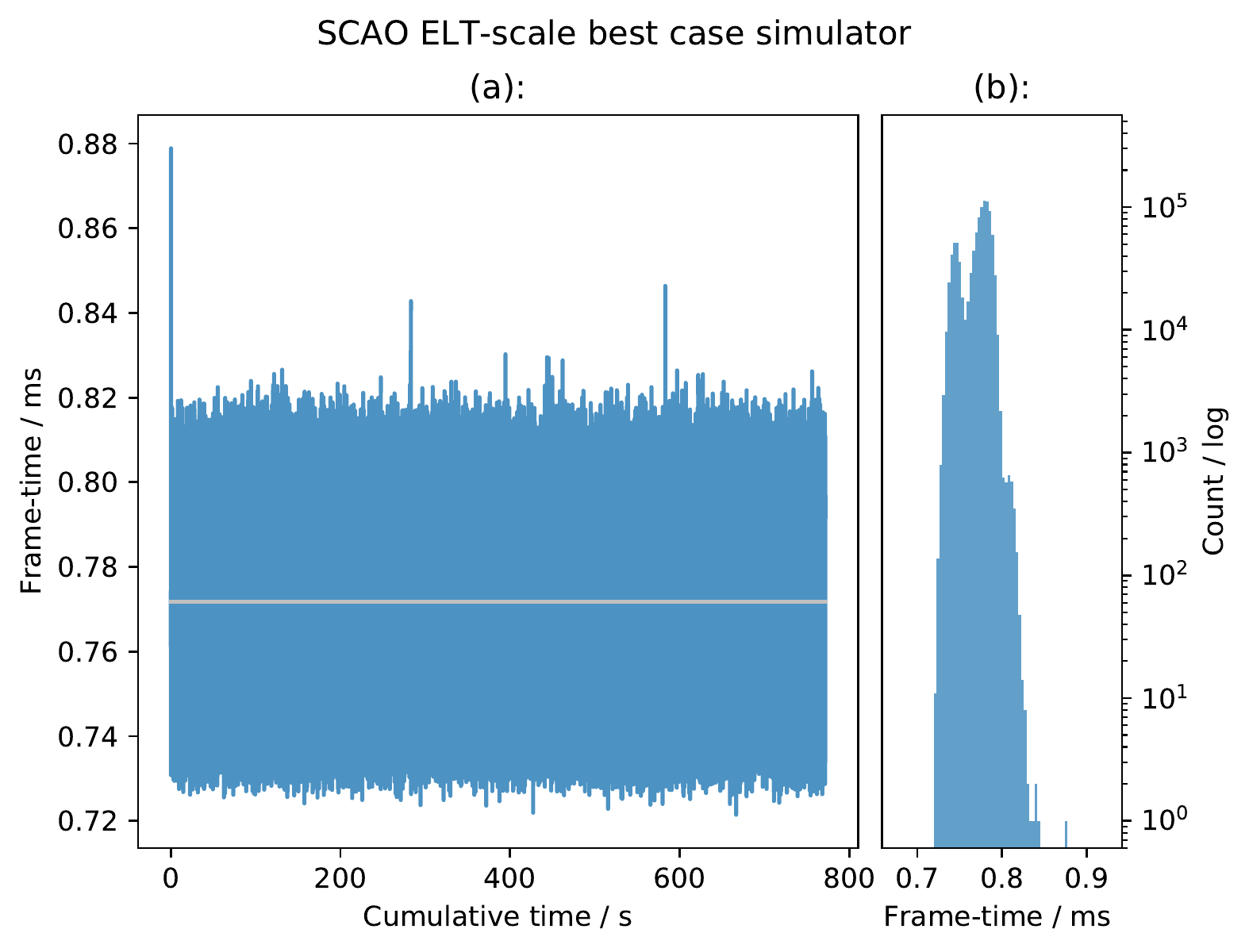}
\caption{a) Frametime results of SCAO best case simulator for \num{e6} frames with an average frametime of \SI{0.77+-0.02}{\milli\second} which corresponds to an average framerate of \SI{1300+-30}{\hertz}. The horizontal line is the average frametime. b) A histogram of the frame times in (a).}
\label{fig:bestcaseframes}
\end{figure}

Figure \ref{fig:bestcaseframes} shows the frametime results of the SCAO best case simulator for \num{e6} frames.  The figure shows the minimal number of outliers and also the small spread of the distribution. The average frametime of \SI{0.77+-0.02}{\milli\second} corresponds to an average framerate of \SI{1300+-30}{\hertz}.  This is shorter than a typical atmospheric coherence time, and therefore would be suitable for ELT-scale SCAO RTC.  The rms jitter is \SI{16.3}{\micro\second}, which is about \SI{2}{\percent} of the mean frame time, and would have insignificant impact on AO performance \citep{Pettazzi2012}.  The maximum instantaneous peak-to-peak jitter between consecutive frames is \SI{107}{\micro\second}, including the startup measurements, or \SI{88.8}{\micro\second} during the long-term measurements.

\subsubsection{DARC on Xeon Phi for ELT scale AO RTC} \label{sec:darcnocam}

\begin{figure}
\includegraphics[width=\columnwidth]{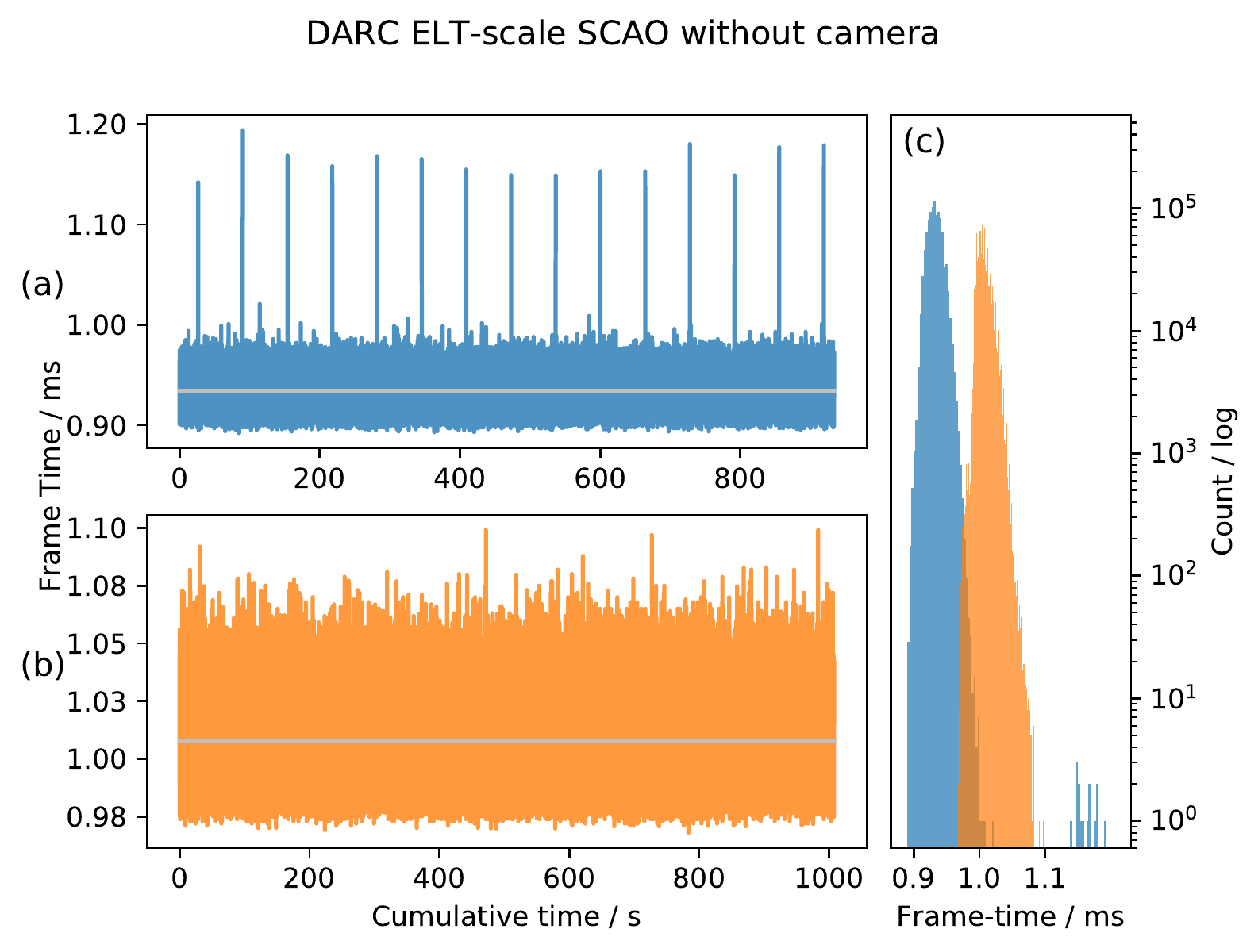}
\caption{a) Frametimes for DARC SCAO on an Intel Motherboard with no camera for $10^6$ frames with an average frametime of \SI{0.93+-0.01}{\milli\second} which corresponds to an average framerate of \SI{1070+-10}{\hertz}. b) As for (a), except for a Supermicro motherboard with an average frametime of \SI{1.01+-0.01}{\milli\second} which corresponds to an average framerate of \SI{990+-10}{\hertz}. c) Histograms of the frametimes presented in (a) and (b) with (a) on the left and (b) on the right. The horizontal lines in (a) and (b) are the mean frametimes for each distribution. }
\label{fig:darcnocamframes}
\end{figure}

Figure \ref{fig:darcnocamframes} shows the frametime results of DARC configured for ELT-scale SCAO (in a similar configuration as above however with 5316 actuators, to mimic the ELTs M4 Adaptive mirror, and 4632 subapertures for a total of 9264 slope measurements) though without a physical camera connected, measured over \num{e6} frames.  It can be seen in Figure~\ref{fig:darcnocamframes}(a) that for the system using an Intel motherboard (model S72000, 7250 processor), there are a small number of regular single frame outliers which add about \SIrange{200}{250}{\micro\second} to the frame time, roughly every \SI{63.75}{\second}.  We have determined that these events are due to the Intel system management interface on the motherboard, which periodically polls the processor for information.  There appears to be no way in which this can be turned off.  The presence of these interrupts can be verified using this code:
\begin{enumerate}
\item for SEC in `seq 0 200`; do echo -n "\$SEC "; rdmsr -p 0 -d 0x34; sleep 1; done,
\end{enumerate}
which has been used to confirm their presence on the Intel S72000 motherboard used in this report.

Figure~\ref{fig:darcnocamframes}(b) shows results taken using a Ninja Development platform Xeon Phi using a Supermicro motherboard (model K1SPE with 7210 processor).  Here it can be seen that these \SI{64}{\second} period events are not present.  It is therefore important to take care when evaluating motherboards suitable for AO RTC. Histograms of both measurements are shown in Figure~\ref{fig:darcnocamframes}(c), the difference in mean frametime between the two distributions is due to the specification of the processors used in each motherboard; \SIrange[range-phrase=~vs.~]{1.4}{1.3}{\giga\hertz} clock speed, \SIrange[range-phrase=~vs.~]{480}{450}{\giga\byte\per\second} memory bandwidth (Table~\ref{tab:knlspec}). From this figure, it can be seen that the distribution of latency measurements is approximately Gaussian, except for the outliers.

Therefore, DARC is able to operate ELT-scale SCAO with a \SI{0.93+-0.01}{\milli\second} frame time, corresponding to a \SI{1070+-10}{\hertz} maximum frame rate.  When the \SI{64}{\second} events are included, the instantaneous peak-to-peak jitter over a million frames is \SI{263}{\micro\second}, while ignoring these events reduces the peak-to-peak jitter to \SI{92.7}{\micro\second} and the RMS jitter is only \SI{11.4}{\micro\second}. The Ninja development platform can operate ELT-scale SCAO with a \SI{1.01+-0.01}{\milli\second} frame time, corresponding to a \SI{990+-10}{\hertz} maximum frame rate. This is a lower maximum performance than the Intel motherboard system due to the difference in processor specification which is as expected.

\subsubsection{Storing the control matrix as \SI{16}{\bit} floating point values} \label{sec:halffloatresults}

For investigating the effect on the latency of the RTC on the Xeon Phi when storing the control matrix as \SI{16}{\bit} floating point values we had to use our own implementation of an MVM algorithm. This is because the Intel MKL library that is used to calculate the reconstruction MVM in previous results is unable to operate directly on \SI{16}{bit} floating point values. To be able to use it, the values would need to be converted to \SI{32}{\bit} before each call to the MKL library. This would not be ideal as MKL works best on larger MVM problem sizes and converting a large amount of the control matrix to \SI{32}{\bit} would defeat the purpose of storing it as \SI{16}{\bit} and making too many calls to the library would vastly increase the latency.

Our custom MVM implementation uses Intel intrinsic instructions to convert the \SI{16}{\bit} values of the control matrix immediately before they are operated upon. Up to 16 conversions can be done in a single instruction with the results stored in one of the \SI{512}{\bit} vector registers ready for the FMA MVM operations. This ensures that there is a minimum transfer of \SI{32}{\bit} values to conserve memory bandwidth.

A similar custom \SI{32}{\bit} MVM implementation, simply loading the data instead of converting it, shows that this algorithm isn't as optimised as MKL. It gives an average frametime of \SI{0.995+-0.006}{\milli\second} with an RMS jitter of \SI{5.96}{\micro\second} which can be compared to results that use MKL on the same processor/motherboard of \SI{0.93+-0.01}{\milli\second} from Figure~\ref{fig:darcnocamframes}.

The algorithm that uses a \SI{16}{\bit} control matrix decreases this average frametime to \SI{0.902+-0.006}{\milli\second} with an RMS jitter of \SI{5.60}{\micro\second}. The need to use a less optimised custom MVM and the need to convert to \SI{32}{\bit} floats introduces extra overhead which greatly reduces the potential gain. These results show that this implementation of storing and converting the 16-bit control matrix increases performance by only \SI{3}{\percent} over the best case MKL results.

The next iteration of Xeon Phi after KNL, Knights Mill (KNM) which is available now, includes support for Intel variable precision operations (vector neural network instruction, VNNI) which include intrinsic instructions that can directly operate on \SI{16}{\bit} integer values by addition and multiplication to produce an accumulated \SI{32}{\bit} integer sum. This would require some conversion of the control matrix and slope values to fixed-point \SI{16}{\bit} precision integers but could reduce the memory bandwidth requirement without introducing a costly \SIrange[range-phrase=~to~]{16}{32}{\bit} conversion for each control matrix value at execution time. Simulations have shown that \SI{16}{\bit} fixed-point values would just be sufficient to provide the required precision \citep{halfPrecisionFloat}, however when taking into account a real system with misalignments it may not be adequate.

This functionality comes via a redesigned vector processing unit (VPU) which also doubles the number of SP FMA operations possible per instruction cycle, increasing the theoretical peak SP-FLOPS 2X over KNL. This is achieved with quad FMA (QFMA) instructions which allow for sequential FMA to accumulate over four sets of calculations with a single instruction. There are caveats to this however due the instruction pipeline of the VPUs, so 2X speed up is unlikely, but the QFMA operations should definitely benefit the highly vectorisable MVM without needing to use the \SI{16}{\bit} VNNI; investigation into KNM VNNI and QFMA instructions could be useful future work.

\subsection{DARC SCAO computation with a real wavefront sensor camera} \label{sec:darcwithcam}

\begin{figure}
\includegraphics[width=\columnwidth]{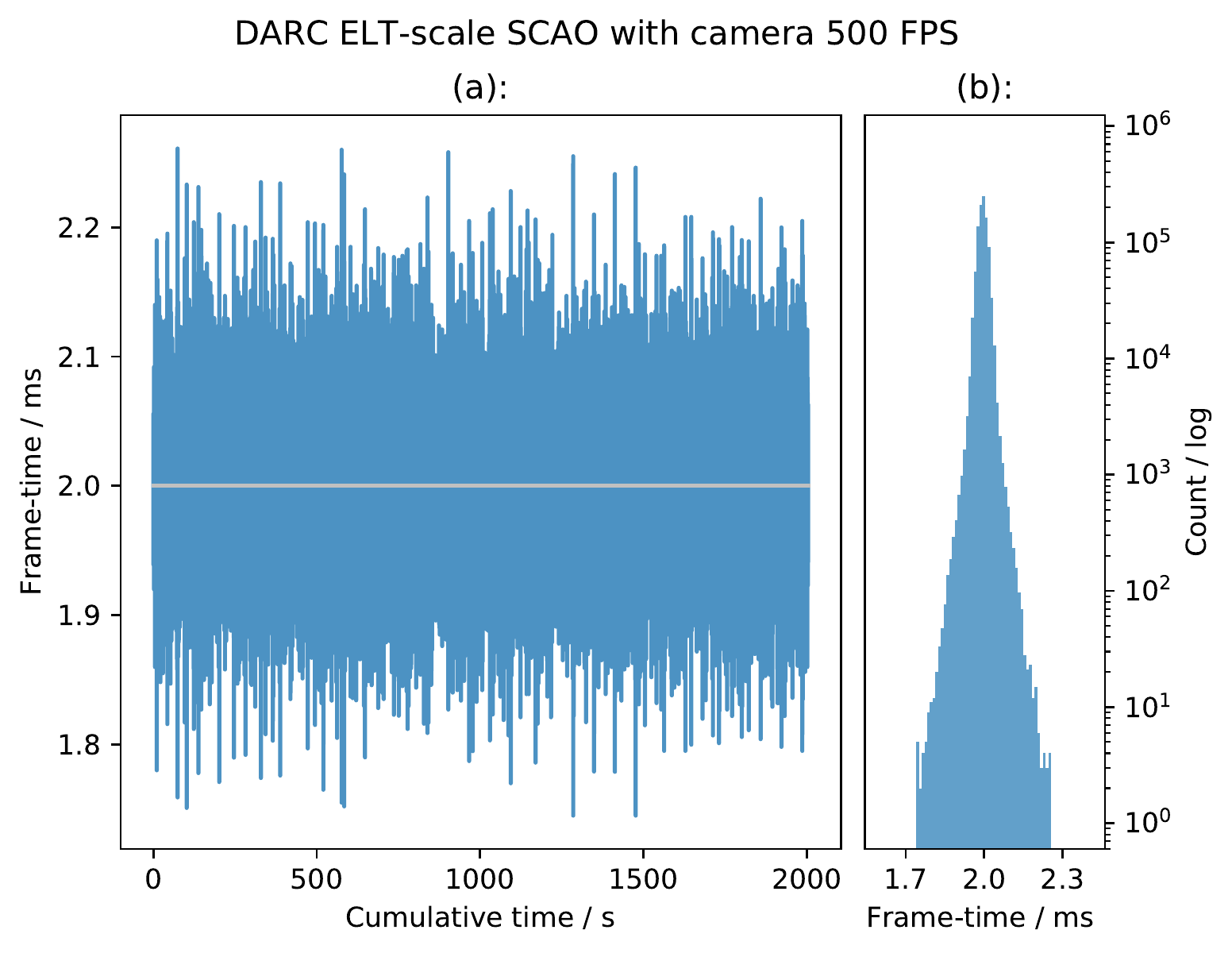}
\caption{a) Frametimes of DARC SCAO with a real wavefront sensor camera operating at \SI{500}{\hertz} for \num{e6} frames, the horizontal line shows the average frametime of \SI{2.00+-0.02}{\milli\second}. b) A histogram of the frame times, showing the distribution of jitter.}
\label{fig:darccamframes500}
\end{figure}

\begin{figure}
\includegraphics[width=\columnwidth]{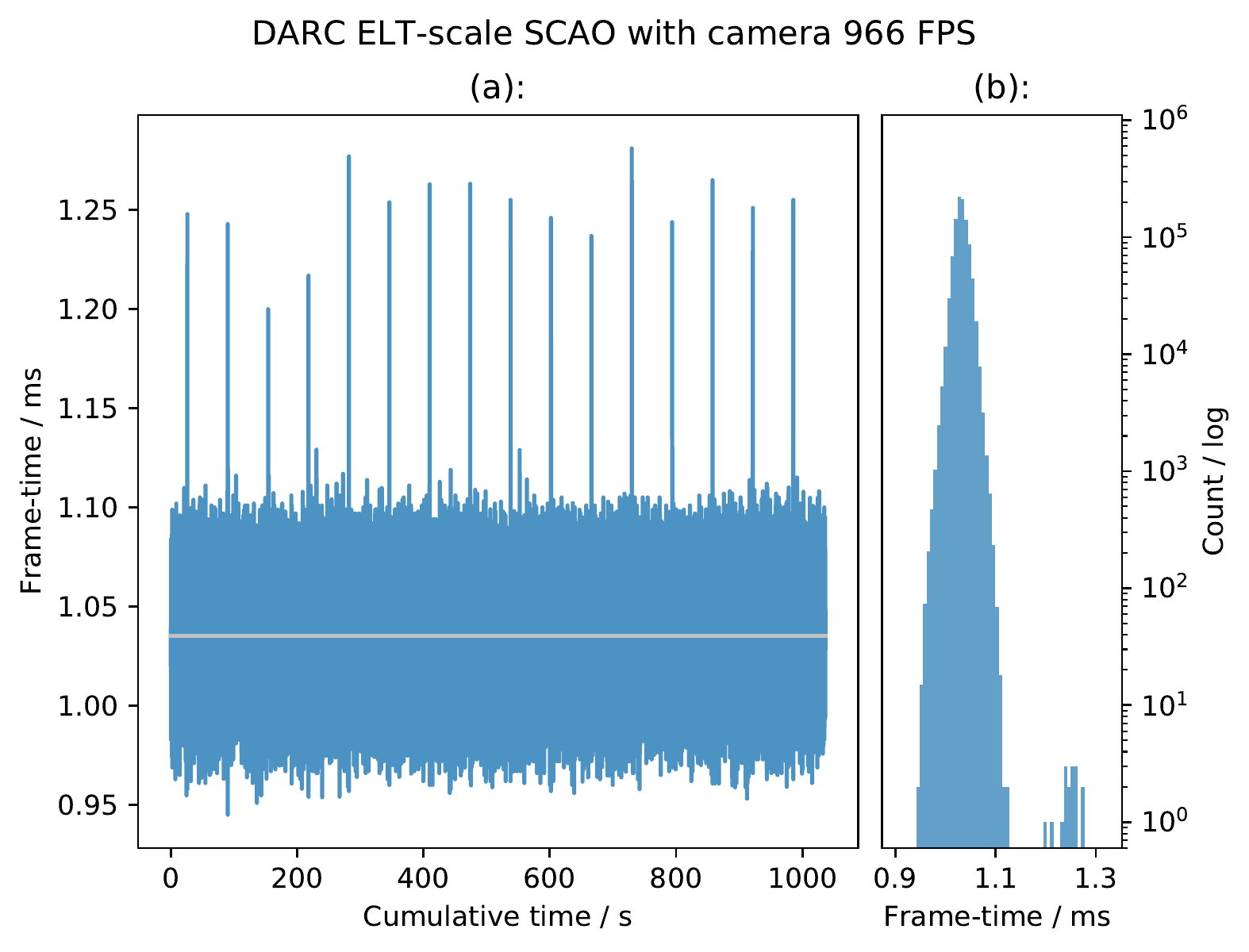}
\caption{a) Frametimes of DARC SCAO with a real wavefront sensor camera operating at \SI{966}{\hertz} for \num{e6} frames, the horizontal line shows the average frametime of \SI{1.04+-0.01}{\milli\second}. b) A histogram of the frame times, showing the distribution of jitter. }
\label{fig:darccamframes966}
\end{figure}

Figure~\ref{fig:darccamframes500} and Figure~\ref{fig:darccamframes966} show two sets of frametime results for DARC configured for ELT-scale SCAO, with pixels arriving from a real 10GigE Vision based camera running at \SI{500}{\hertz} and at the camera's maximum framerate of \SI{966}{\hertz} respectively. The camera is an Emergent Vision Technologies HS2000M, delivering 100 pixels per sub-aperture.  The figures show minimal numbers of outliers and also the small spread of the distributions, with rms jitters of \SI{20.1}{\micro\second} and \SI{13.8}{\micro\second} for \SI{500}{\hertz} and \SI{966}{\hertz} respectively, which is similar to that when DARC operates without a real camera.

The \SI{64}{\second} events due to the Intel motherboard are visible in the data for \SI{966}{\hertz}, however they are not seen in the data for \SI{500}{\hertz}, this likely due to the reduced computational demands for SCAO at \SI{500}{\hertz} and so the CPU has ample time to process the interrupts without affecting DARC. The maximum instantaneous peak-to-peak jitter is \SI{510}{\micro\second} for \SI{500}{\hertz} and \SI{163}{\micro\second} for \SI{966}{\hertz} excluding the \SI{64}{\second} events, over one million frames.

The shapes of the histograms in Figure~\ref{fig:darccamframes500} and Figure~\ref{fig:darccamframes966} are quite different, as they are plotted on a log scale it shows that for the camera operating at \SI{500}{\hertz} there is a high narrow peak at the mean of the distribution with relatively small numbers of frames spread out to either side. This gives the distribution it's low RMS jitter but a relatively high relative instantaneous peak-to-peak jitter.

We use a modified version of the Aravis GigE Vision Library \citep{Aravis}, which enables access to the pixel stream, rather than waiting until the entire frame has been delivered. In this way, DARC can begin processing sub-apertures as soon as enough pixels have arrived, reducing latency. As the latency of an RTC is defined as the time between last pixel arriving and the final DM command being sent out, reducing this time improves the performance of the RTC.

\begin{figure}
    \includegraphics[width=\columnwidth]{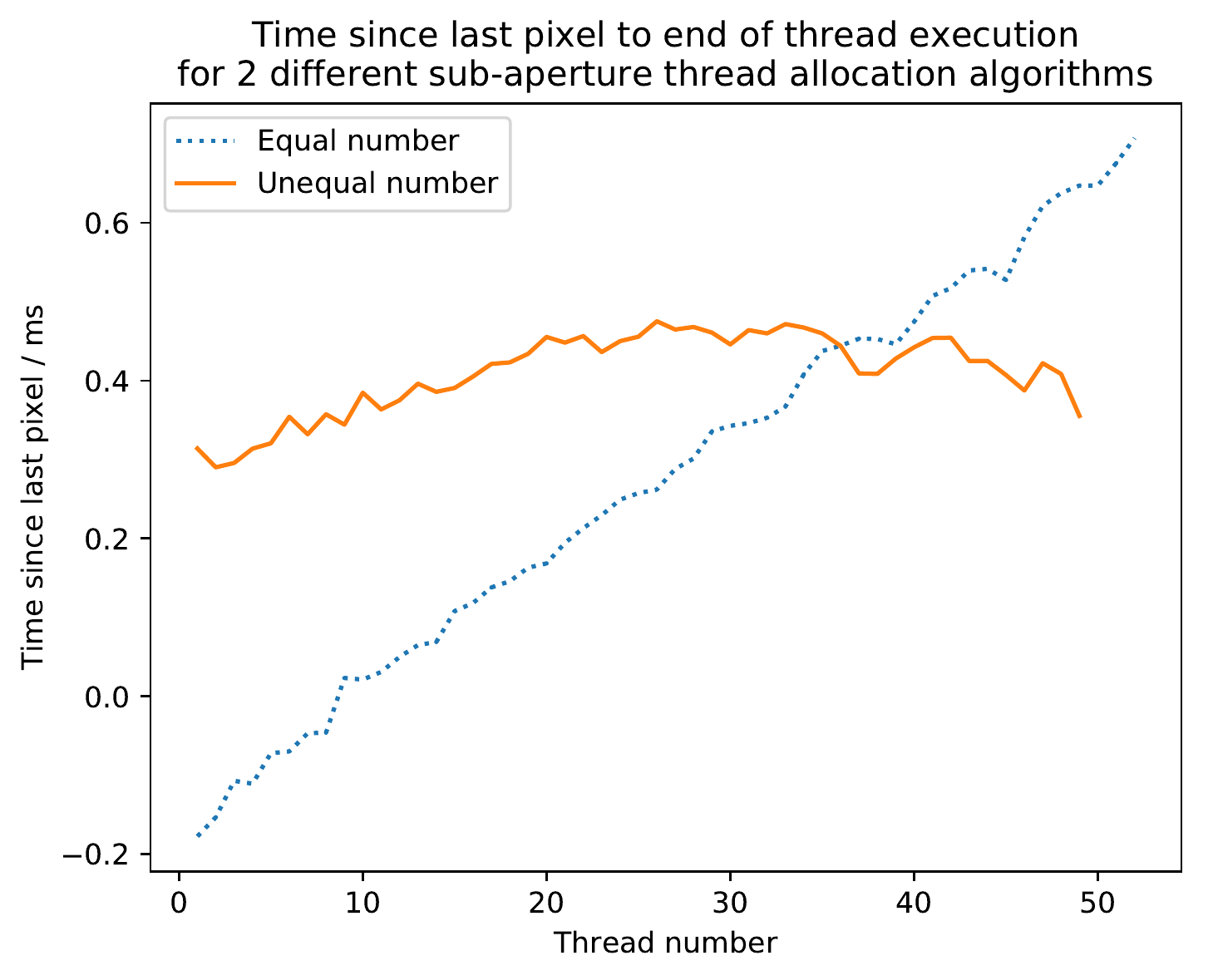}
    \caption{A measure of the time from the last pixel arriving from the camera to end of thread computation per thread. `Equal number' shows the results for a naive subaperture allocation whereby each thread processes an equal number of subapertures. `Unequal number' shows allocation by a simple algorithm which gives more work to threads which are processing subapertures whose pixels arrive earlier.}
    \label{fig:threadtimes}
    \end{figure}

\begin{figure}
\includegraphics[width=\columnwidth]{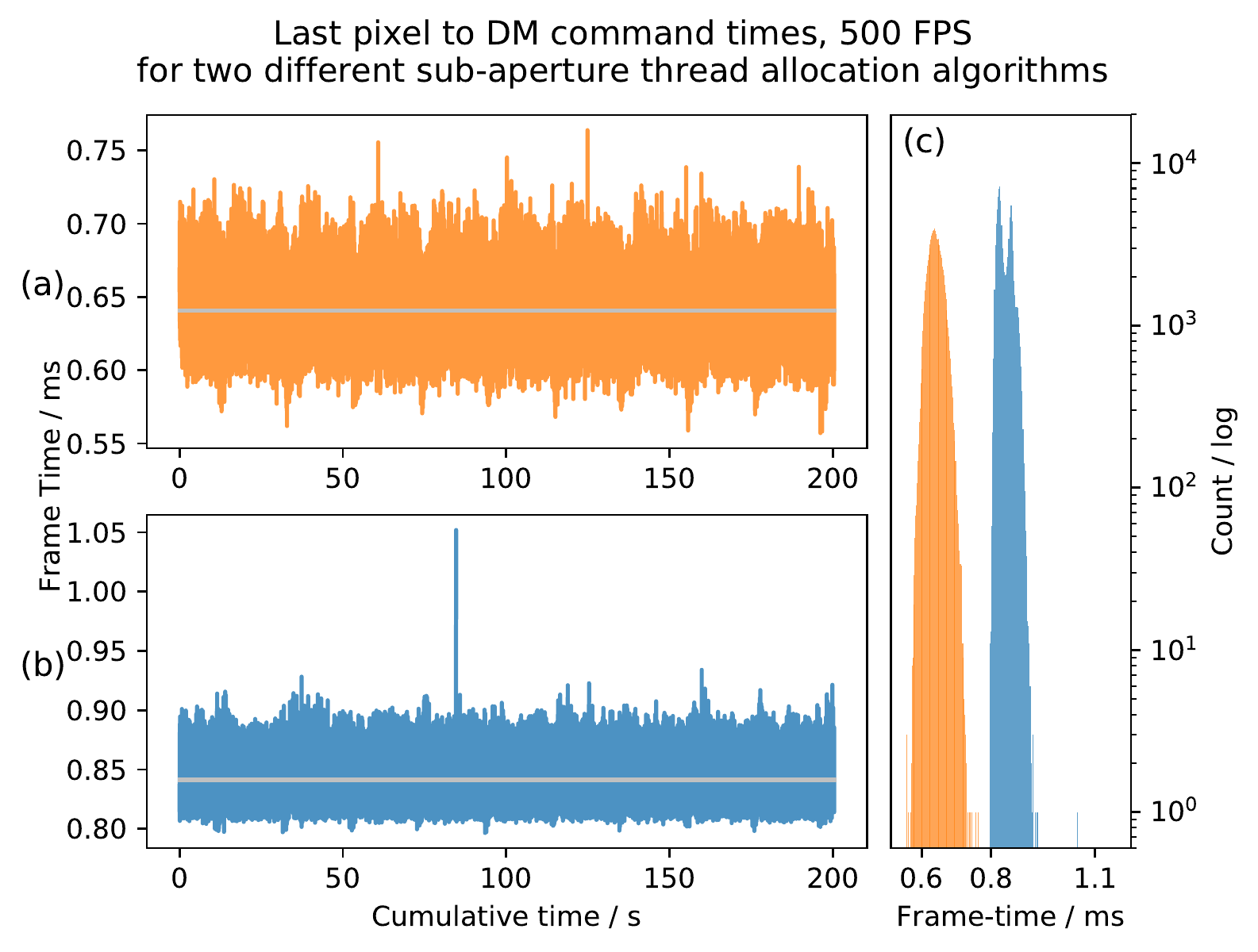}
\caption{Latency measurements (time between last pixel received to DM demand ready) for DARC operation with a real wavefront sensor camera at ELT-scale at \SI{500}{\hertz}. a) shows results when using the unequal subaperture thread allocation. b) shows results for the equal subaperture thread allocation, Figure~\ref{fig:threadtimes}.}
\label{fig:darccamlatency}
\end{figure}

Figure~\ref{fig:threadtimes} shows the time taken for the DARC processing threads to finish processing their subapertures from the time the last pixel arrives from a real camera; these are average times for \num{e5} frames. Figure~\ref{fig:threadtimes} (Equal number) is for the case described in Figure~\ref{fig:pipelining}(c) and Figure~\ref{fig:threadtimes} (Unequal number) is for the Figure~\ref{fig:pipelining}(d) case. Both sets of data are taken with the real camera operating at \SI{500}{\hertz}. Figure~\ref{fig:darccamlatency} shows that for the situation with equal numbers of subapertures the mean RTC latency for \num{e5} frames is \SI{0.84+-0.02}{\milli\second}, and for unequal numbers, the mean RTC latency is \SI{0.64+-0.02}{\milli\second}. Figure~\ref{fig:threadtimes} shows a modest improvement in RTC latency, bringing the latency below that of the best case simulator and demonstrates the different end of thread execution times described in Figure~\ref{fig:pipelining}. These results show that DARC on the Xeon Phi can operate SCAO at ELT-scales with a real camera.

The algorithm used to assign the unequal numbers of subapertures is a very basic implementation with a linearly decreasing sub-aperture count per thread. This algorithm will be explored further to find the optimal sub-aperture allocation to improve latency for desired framerates and different read-out rates.

\subsection{Best-case and DARC performance}
The results described in Sections \ref{sec:darcnocam} and \ref{sec:darcwithcam} for the mature DARC AO RTC on the Xeon Phi are consistent with those found in Section \ref{sec:bestcasescao}. The frametimes for DARC with no camera, as shown in Figure \ref{fig:darcnocamframes}, give the current best performance of the computation along with pipelining and  hard-real-time aspects which are required for a real RTC.  DARC operates with a maximum frame rate which is slightly below that of the best case system, though will also exhibit lower latency, due to the ability to interleave processing and pixel acquisition (which the base-case system cannot do). Figure~\ref{fig:darccamlatency} shows that by using a more appropriate subaperture thread allocation algorithm, the latency of the RTC with a real camera can be reduced to well below that of the best case simulator.

\subsection{Future Work}
The optimisations of DARC presented in this paper are far from exhaustive, they focus on the multi-threading and reconstruction aspects and provide us with adequate performance for ELT SCAO. However, they can still be improved upon by further investigation into other aspects of the real-time pipeline such as the pixel calibration and centroiding parts. There are also more novel algorithms in use in AO which could benefit from acceleration by Xeon Phi such as more complex reconstruction techniques like Linear Quadratic Gaussian \citep[LQG,][]{Kulcsar2006} reconstruction and less computationally expensive techniques such as the Cumulative Reconstructor with domain decomposition \citep[CuRe-D,][]{Rosensteiner2012}.

Another area where Xeon Phis may be applicable is for ELT scale AO modelling and simulation, to aid design decisions during system design studies. Currently full ELT scale simulations are not capable of being run even close to real-time \citep{Basden2014}, and therefore the large AO parameter space takes a long time to fully explore.  To aid with ELT commissioning, a hardware-in-the-loop real-time simulation \citep{Basden2014} will be required, and here, operation at close to real-time rates becomes very important.  AO PSF reconstruction will also benefit from real-time simulation rates.

Investigation into the improved \SI{16}{\bit} integer instructions, such as VNNI and QFMA described in Section~\ref{sec:halffloatresults}, provided by the Xeon Phi Knights Mill could also be considered for future investigation. Hardware that supports half-precision integer instructions are becoming ever more common with their use in neural network applications and so their investigation for AO RTC would be very useful for future hardware.

\subsubsection{ELT MCAO prototype}
This paper has only discussed the SCAO regime of AO for ELT-scale but the next generation telescopes all have more complex multi-conjugate or multi-object AO systems planned for which suitable RTCs will be required. One of these systems is the ELTs Multi-conjugate Adaptive Optics Relay (MAORY) which will employ 6 Laser Guide Stars (LGS) along with three Natural Guide Stars (NGS) using 3 DMs for the correction, increasing the computation requirements by more than 6 times that of ELT SCAO. Our initial investigation \citep{Jenkins2018} for this involves multiple Xeon Phi systems working in parallel to process the images from each WFS independently before combining the partial DM vectors and delivering to the DMs. This presents significant challenges in addition to those for preparing DARC for ELT-scale SCAO.

\section{Conclusion} \label{sec:conclusion}
The results presented show that the Intel Xeon Phi is capable of performing the computational requirements of a full on-sky tested AO RTC for ELT scale with frame-rates and latencies which are suitable for AO system operation. Furthermore, it has been shown that with a real wavefront sensor camera, the RTC is capable of performing at similar frame-rates albeit with slightly increased jitter.  We can recommend the Xeon Phi as suitable hardware for ELT-scale real-time control, primarily due to the large number of cores, and the high memory bandwidth, though note that it is necessary to take care when selecting a suitable motherboard, to avoid periodic temporary latency increases.